\documentclass[conference]{IEEEtran}
\IEEEoverridecommandlockouts
\usepackage{cite}
\usepackage{amsmath,amssymb,amsfonts}
\usepackage{algorithmic}
\usepackage{graphicx}
\usepackage{textcomp}
\usepackage{xcolor}
\usepackage{subcaption}
\usepackage{booktabs}
\usepackage{bm}
\usepackage{multirow}

\def\BibTeX{{\rm B\kern-.05em{\sc i\kern-.025em b}\kern-.08em
    T\kern-.1667em\lower.7ex\hbox{E}\kern-.125emX}}
\begin{document}

\title{Performance of Graph Neural Networks for Point Cloud Applications
}

\author{
\IEEEauthorblockN{ Dhruv Parikh\IEEEauthorrefmark{1}, Bingyi Zhang\IEEEauthorrefmark{1}, Rajgopal Kannan\IEEEauthorrefmark{2}, Viktor Prasanna\IEEEauthorrefmark{1}, Carl Busart\IEEEauthorrefmark{2}}
\IEEEauthorblockA{
    \IEEEauthorrefmark{1}University of Southern California \IEEEauthorrefmark{2}DEVCOM US Army Research Lab\\
    \IEEEauthorrefmark{1}\{dhruvash, bingyizh, prasanna\}@usc.edu \IEEEauthorrefmark{2}\{rajgopal.kannan.civ, carl.e.busart.civ\}@army.mil}
}

\maketitle

\begin{abstract}


Graph Neural Networks (GNNs) have gained significant momentum recently due to their capability to learn on unstructured graph data. Dynamic GNNs (DGNNs) are the current state-of-the-art for point cloud applications; such applications (viz. autonomous driving) require real-time processing at the edge with tight latency and memory constraints. Conducting performance analysis on such DGNNs, thus, becomes a crucial task to evaluate network suitability. 


This paper presents a profiling analysis of EdgeConv-based DGNNs applied to point cloud inputs. We assess their inference performance in terms of end-to-end latency and memory consumption on state-of-the-art CPU and GPU platforms. The EdgeConv layer has two stages: (1) dynamic graph generation using $k$-Nearest Neighbors ($k$NN) and, (2) node feature updation. The addition of dynamic graph generation via $k$NN in each (EdgeConv) layer enhances network performance compared to networks that work with the same static graph in each layer; such performance enhancement comes, however, at the added computational cost associated with the dynamic graph generation stage (via $k$NN algorithm). Understanding its costs is essential for identifying the performance bottleneck and exploring potential avenues for hardware acceleration. To this end, this paper aims to shed light on the performance characteristics of EdgeConv-based DGNNs for point cloud inputs. Our performance analysis on a state-of-the-art EdgeConv network for classification shows that the dynamic graph construction via $k$NN takes up upwards of $95\%$ of network latency on the GPU and almost $90\%$ on the CPU. Moreover, we propose a quasi-Dynamic Graph Neural Network (qDGNN) that halts dynamic graph updates after a specific depth within the network to significantly reduce the latency on both CPU and GPU whilst matching the original networks inference accuracy.  
\end{abstract}

\begin{IEEEkeywords}
Graph neural network, point cloud, $k$-nearest neighbors, dynamic graph construction, performance profiling
\end{IEEEkeywords}

\section{Introduction}
\label{sec:introduction}

\IEEEPARstart{G}{raphs} are effective data structures for representing intricate relationships (edges) among entities (nodes) with a high degree of interpretability. This has led to the widespread adoption of graph theory in various domains. 

Notably, Graph Neural Networks (GNNs) \cite{kipf2017semisupervised} have demonstrated remarkable success in addressing both conventional tasks like computer vision \cite{garcia2017few, qi20173d} and natural language processing \cite{peng2018large, marcheggiani2017encoding}, as well as non-traditional tasks such as protein interface prediction \cite{fout2017protein} and combinatorial optimization \cite{khalil2017learning}. This versatility has made GNNs an integral part of deep learning methodologies, as evidenced by their numerous applications across diverse problem domains \cite{kipf2017semisupervised, ZHOU202057}.

A point cloud is an unstructured collection of raw data points, where each point represents a specific location associated with an object or shape, typically defined within a coordinate system (such as Cartesian or spherical). The ability of a point cloud to capture 3D environments make it crucial for numerous scene understanding tasks \cite{Jaritz_2019_ICCV, Hou_2021_CVPR} and applications across various domains. For instance, point clouds play a vital role in autonomous driving \cite{9173706, 9380166}, virtual reality \cite{alexiou2020pointxr}, augmented reality \cite{chen2019overview}, construction industry \cite{wang2019applications}, robotics, computer graphics, and many more. The availability of affordable and accessible point cloud acquisition systems, such as LIDAR scanners and RGBD cameras \cite{6162882}, has further emphasized the importance of efficient learning and processing techniques for point clouds.

Prior to the advent of deep learning-based methods, learning on point clouds involved constructing hand-crafted features \cite{han20203d}. With the introduction of deep learning, the techniques applied to point clouds can be broadly classified into two categories. These two categories are differentiated based on the pre-processing steps performed on the point clouds prior to network input \cite{rs12111729}:
(1) \emph{Structured-grid based networks:} These networks preprocess point clouds into a structured input format for the deep neural network. Structuring is typically achieved by generating representative views from the point clouds \cite{su2015multiview, Meyer2019LaserNetAE, Yang_2018_CVPR} or by voxelizing the point clouds into 3D grids \cite{7353481}.
(2) \emph{Raw point cloud based networks:} In this category, networks operate directly on the raw point clouds with minimal to no preprocessing. One approach involves passing multi-layer perceptrons (MLPs) through individual point features to learn local spatial relationships \cite{qi2017pointnet, qi2017pointnetplusplus, li2018pointcnn}. Another approach involves systematically constructing graphs using point features and applying Graph Neural Networks (GNNs) to learn labels \cite{wang2019dynamic, landrieu2018largescale, 9010020}.


Structured-grid based networks suffer from drawbacks associated with point cloud transformation. Transforming point clouds into voxels or image views can be computationally expensive and result in bulky data that is difficult to handle. Moreover, these transformations introduce various errors, such as quantization errors, which can impact the inherent properties of point clouds \cite{qi2017pointnet}.
To address these issues, raw point cloud based networks have emerged as a solution. Among these networks, EdgeConv based networks \cite{wang2019dynamic, wang2021object, zhang2019linkeddgcnn} have become state-of-the-art in various point cloud-related tasks. They are an advancement over the basic PointNet \cite{qi2017pointnet} style architectures that directly operate on point features. The EdgeConv layer excels in learning local features and relationships within point clouds while maintaining permutation invariance \cite{pmlr-v119-garg20c}. The dynamic graph construction in the learned feature space allows EdgeConv to capture semantic similarities between points, regardless of their geometric separation.
By leveraging these strengths, EdgeConv-based networks have shown remarkable performance in tasks involving point clouds, overcoming the limitations associated with structured-grid based approaches.

Point cloud processing finds significant application in autonomous vehicles, which serves as a prominent use case for edge computing. These applications operate under stringent latency and memory constraints \cite{8744265}. In intelligent visual systems within autonomous vehicles, point cloud processing typically constitutes just one stage within a multi-stage compute pipeline. As a result, the imposed constraints on latency and memory become even more critical and imperative to address \cite{grigorescu2020survey}.

Analyzing and profiling the performance of networks that process point clouds is a crucial task, particularly considering the prominence of EdgeConv-based networks in this domain. In this work, we make the following contributions:

\begin{itemize}
    \item  \textbf{Latency analysis:} We perform an in-depth analysis of end-to-end layer-wise latency for EdgeConv-based networks used in classification and segmentation tasks on both state-of-the-art CPU and GPU platforms.
    \item \textbf{Breakdown analysis:} Given the two-stage operation of the EdgeConv layer involving (1) dynamic graph construction and (2) GNN-based node feature updation, we perform  breakdown analysis between $k$NN graph construction and GNN-based feature updating at each layer and across different layers in EdgeConv networks.
    \item \textbf{Effects of varying $k$:} EdgeConv networks dynamically construct a graph from a point cloud using the $k$-nearest neighbors ($k$NN) algorithm. We study the effects of varying the value of $k$ from its optimal value, which is determined through a validation set. This analysis examines the impact of varying $k$ on the network's inference accuracy, inference latency, and memory consumption.
    \item  \textbf{Quasi-Dynamic GNNs:} Dynamic graph construction after each layer in a Dynamic GNN (DGNN) improves its performance compared to static GNN counterparts. We investigate the extent to which performance improvement is affected when employing a quasi-Dynamic strategy, where the graph is made static towards the end of the network.
    \item \textbf{Memory consumption:} We analyze the memory consumption of EdgeConv networks on both CPU and GPU platforms.
    \item \textbf{Bottleneck analysis:} By identifying performance bottleneck in the aforementioned networks, we suggest potential research directions that could help mitigate such bottlenecks and improve overall performance.
    \item  \textbf{Hardware acceleration opportunities:} We discuss the potential for hardware acceleration of EdgeConv-based networks on FPGA devices, which has garnered significant research interest due to its promising benefits.
\end{itemize}

Our aim is to deepen our understanding of EdgeConv-based networks' performance characteristics when processing point clouds and provide insights into optimization opportunities and hardware acceleration possibilities.

Section \ref{sec:background} briefly introduces GNNs and GNNs as they relate to point clouds. Section \ref{sec:experimental-setting} describes the networks and datasets used for the experiments, along with the computing platforms on which the experiments were performed. Results are analyzed in section \ref{sec:result-and-analysis}. Discussion and conclusion follow in section \ref{sec:discussion} and \ref{sec:conclusions} respectively.







\section{Background}
\label{sec:background}

\subsection{Graph Neural Networks (GNNs)}
\label{subsec:GNN}

Graphs $\mathcal{G} = (\mathcal{V}, \mathcal{E})$ are defined via a set of vertices (nodes) $\mathcal{V}$ and a set of edges (links) connecting these vertices, $\mathcal{E} = \{(j, i): j, i \in \mathcal{V}$ and a link exists from $j \rightarrow i$\}. Edges typically have directional context, symbolized via the ordered pairs; an undirected edge between nodes $i$ and $j$ is represented often via two directed edges: $(i, j)$ and $(j, i)$. $\bm{A} \in \mathbb{R}^{n \times n}$ is the adjacency matrix for a graph with $n$ nodes representing the above edges and their weights (for a weighted graph).

A graph can have features associated with both its nodes and edges - additionally, a graph may even have global graph level attributes. Typical learning tasks on graphs occur at either the node level (viz. node classification) \cite{rong2020dropedge}, edge level (viz. link prediction) \cite{zhang2018link} or at the graph level \cite{errica2019fair}.


Neural message passing is a mechanism in GNNs whereby information is passed (shared) across nodes and edges within the graph to update node embeddings via a set of learnable parameters \cite{gilmer2017neural}. GNNs employing this message passing framework are called Message Passing Neural Networks (MPNNs) \cite{gilmer2017neural}. 

\begin{figure}[htbp]
\centerline{\includegraphics[width = \linewidth]{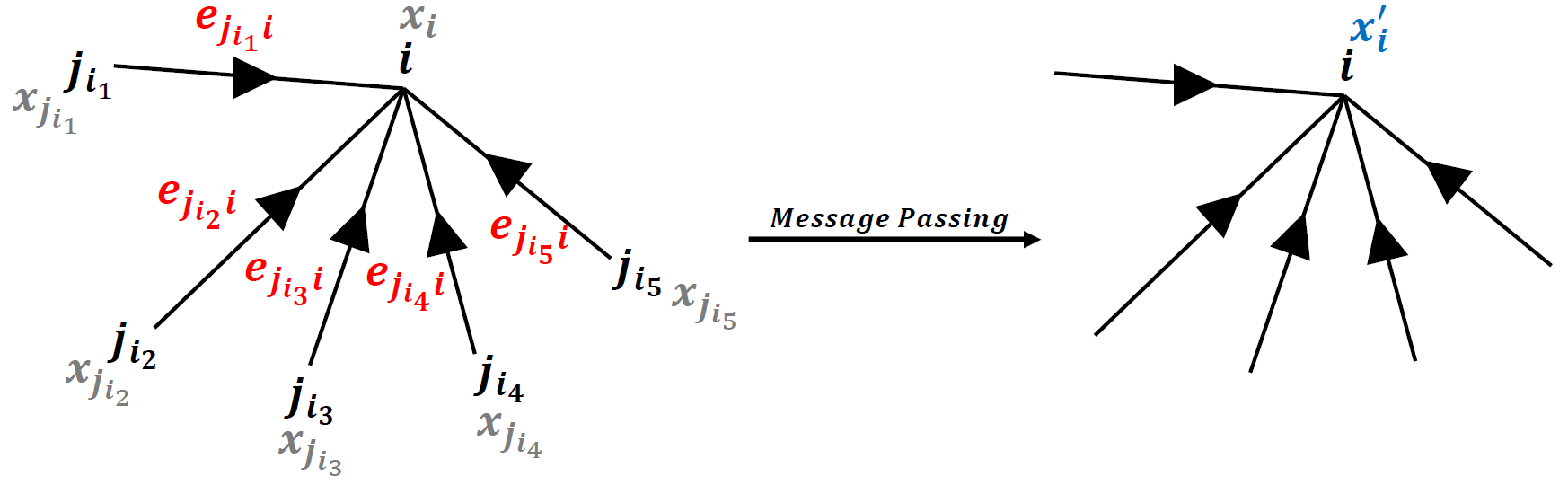}}
\caption{The message passing mechanism employed in GNNs.}
\label{fig:message-passing}
\end{figure}

For a node $i$ with a node feature vector $\bm{x_i}$ and a neighbor set $\mathcal{N}(i)$, the general formulation for message passing can be described by Equation \ref{eq:message-passing} \cite{gilmer2017neural} (illustrated in Figure \ref{fig:message-passing}),

\begin{equation}
\label{eq:message-passing}
\bm{x_{i}^{'}} = \Psi_{\Theta}(\bm{x_{i}}, \sum_{j \in \mathcal{N}(i)} \mathcal{M}_{\Phi}
(\bm{x_{i}}, \bm{x_{j}}, \bm{e_{ji}})).
\end{equation}
The above equation can be decomposed into the following stages:
\begin{itemize}
    \item \textbf{Message generation.} A message is constructed between a node $i$ and its neighbor $j$ using node level features $\bm{x_i}$ and $\bm{x_j}$ and edge level feature $\bm{e_{ji}}$ as $\mathcal{M}_{\Phi}(\bm{x_{i}}, \bm{x_{j}}, \bm{e_{ji}})$.
    \item \textbf{Aggregation.} Such constructed messages from all of the nodes neighbors are aggregated via the aggregation function $\sum_{j \in \mathcal{N}(i)}$; the aggregation function $\sum$ is typically permutation invariant for point cloud applications.
    \item \textbf{Updation.} Finally, the aggregated message along with $\bm{x_i}$ is used to learn the feature update for node $i$, $\bm{x_{i}{'}}$ via the function $\Psi_{\Theta}$.
\end{itemize}
For a given layer, such message-aggregate-update paradigm is applied to all the nodes of the graph with parameters $(\Theta, \Phi)$ shared across all the nodes within a layer. The functions $\mathcal{M}_{\Phi}$ and $\Psi_{\Theta}$ are typically multi-layer perceptrons ($\mathcal{MLP}s$).

Such GNN layers, employing the message passing framework to update the node features, as above, are central to GNNs such as Graph Convolutional Network (GCN) \cite{kipf2017semisupervised}, GraphSAGE \cite{hamilton2018inductive}, Graph Isomorphism Network (GIN) \cite{xu2019powerful}, Graph Attention Network (GAN) \cite{veličković2018graph}, Principal Neighborhood Aggregation (PNA) \cite{corso2020principal}, etc.

\begin{figure*}
    \centering
    \includegraphics[width=\textwidth]{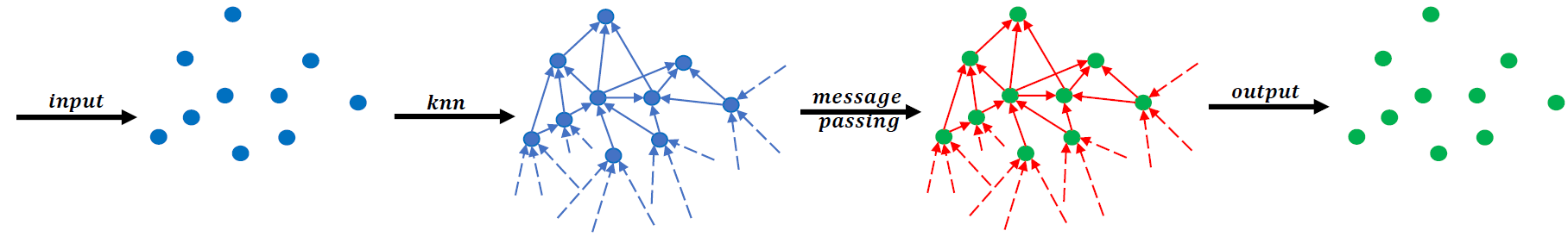}
    \caption{Illustration of the EdgeConv layer. A point cloud (blue) is the input to the EdgeConv layer. EdgeConv layer uses $k$NN to generate a directed $k$NN graph. The message passing paradigm, applied to this graph, uses a nodes neighbors to update node embeddings (message passing: red lines, node feature updates: green dots). The output of the EdgeConv layer is the point cloud with updated node features (green).}
    \label{fig_2}
\end{figure*}

EdgeConv layer \cite{wang2019dynamic} also utilizes this message passing paradigm - however, unlike the above networks, each EdgeConv layer first performs dynamic graph construction via a $k$-nearest neighbor ($k$NN) algorithm to construct a $k$NN graph; a $k$NN graph is a directed graph in which each node is connected to its $k$ nearest neighboring nodes via edges (Figure \ref{fig_2}). Then, the EdgeConv layer performs message passing within this $k$NN graph to update the node embeddings.

\subsection{GNNs for Point Clouds}
\label{subsec:GNN}

The authors in PointNet \cite{qi2017pointnet} introduced the notion of ingesting raw, unordered point clouds to learn directly on the associated point level features. This network, at its core, passes each point level feature vector through a shared $\mathcal{MLP}$ to update the point features. In the message passing framework, 
\begin{equation}
\label{eq:pointnet_message_passing}
\bm{x_{i}{'}} = \Psi_{\Theta}(\bm{x_i})
\end{equation}
Essentially, no messages are passed and no aggregations occur - it directly updates all the node embeddings. The final layer is a global-level max pool layer which generates a global feature vector for graph level tasks. Due to the shared $\mathcal{MLP}$ and global max pool layer, such a network is fully invariant to permutations in the ordering of the $n$ input points. This permutation invariance is a key characteristic of GNNs despite the PointNet not traditionally being a graph neural network.

A class of PointNet derivative networks \cite{atzmon2018point, ravanbakhsh2017deep, zaheer2018deep, klokov2017escape} use associated approaches to learn directly from the point features.
EdgeConv layer uses both message passing and dynamic graph construction, as shown in Figure \ref{fig_2}, to reach state-of-the-art results for point cloud applications. The message passing paradigm in EdgeConv can be described as below, by (3),

\begin{equation}
 \bm{x_{i}^{'}} = \sum_{j \in \mathcal{N}(i)} \mathcal{M}_{\Phi}
(\bm{x_{i}}, \bm{x_{j}} - \bm{x_{i}})
\label{eq:edge_conv_message_passing}
\end{equation}

\begin{equation}
\sum \rightarrow \max(.) 
\label{eq:max}
\end{equation}

\begin{equation}
\mathcal{M}_{\Phi}(\bm{x}, \bm{y}) \rightarrow \mathcal{MLP}_{\Phi}(\bm{x}\,||\,\bm{y})
\label{eq:concatenation}
\end{equation}
where the $\max(.)$ in (\ref{eq:max}) is channel-wise along the nodes and the $||$ in (\ref{eq:concatenation}) is a concatenation operation; $\mathcal{MLP}_{\Phi}$ is a multi-layer perceptron (parameterized by $\Phi$) with a ReLU non-linearity.

The inclusion of the edge feature $\bm{x_j} - \bm{x_i}$ adds local context to the updated node embedding, and the node-level feature $\bm{x_i}$ helps the network retain global information.
A single EdgeConv layer takes in an input tensor $\bm{X} \in \mathbb{R}^{n \times c}$ for a point cloud with $n$ points, each point represented by a vector node embedding of length $c$. The output of $k$NN graph construction on $\bm{X}$ is a tensor $\bm{X'} \in \mathbb{R}^{n \times k \times c}$ representing for each node, its $k$ neighboring nodes and their node embeddings (feature vectors). Before passing $\bm{X'}$ through an $\mathcal{MLP}$ layer, a nodes feature vector is subtracted from the feature vector of its neighbors ($\bm{x_{j}} - \bm{x_{i}}$) and concatenated to the resultant ($\bm{x_i} \, || \, \bm{x_j - x_i}$). The tensor thus obtained, $\bm{X_{in}^{\mathcal{MLP}}} \in \mathbb{R}^{n \times k \times 2c}$ is passed through $\mathcal{MLP}_{\Phi}\{2c,\,a_1,\,a_2,\,a_3,\,...,\,a_m\}$, which is an $m$-layer $\mathcal{MLP}$ with ReLU activations, to generate $\bm{X_{out}^{\mathcal{MLP}}} \in \mathbb{R}^{n \times k \times a_m}$ which is finally aggregated (via $\max$) to generate $\bm{Y} \in \mathbb{R}^{n \times a_m}$ which is the final output from the EdgeConv layer. $\mathcal{MLP}_{\Phi}\{1024, 512, 256\}$ is a $2$-layer perceptron; the input channels to the first layer is $1024$, the output channels of the first layer is $512$ and the output channels of the final (second) layer is $256$.







\section{Experimental Setting}
\label{sec:experimental-setting}

\subsection{Platform Details}
\label{subsec:platform}
The performance analysis for EdgeConv is conducted on state-of-the-art GPU and CPU platforms (See Table \ref{tab:platform-specifications}). The GPU used, for both training and inference, is NVIDIA RTX A6000, which has $10,752$ NVIDIA Ampere architecture CUDA cores. The CPU utilized for inference is AMD Ryzen Threadripper 3990x with 64 CPU cores. Additionally, we use the PyTorch \cite{pytorch} and PyTorch Geometric \cite{PyG} libraries to facilitate training and inference on the above platforms - no additional kernel-level optimization is performed.

\begin{table}[!ht]
\centering
\caption{Specifications of platforms }
\begin{tabular}{c|cccccc}
 \toprule
\textbf{Platforms} & CPU & GPU   \\ 
\midrule \midrule 
Platform  & AMD Threadripper 3990x & Nvidia RTX A6000  \\
 {Platform Technology}  & TSMC 7 nm   & TSMC 7 nm \\ 
{Frequency} & 2.90 GHz  & 1.8 GHz  \\ 
{Peak Performance}& 3.7 TFLOPS & 38.7 TFLOPS  \\ 
{On-chip Memory}& 256 MB L3 cache & 6 MB L2 cache   \\
{Memory Bandwidth}& 107 GB/s & 768 GB/s   \\ \bottomrule
\end{tabular}
\label{tab:platform-specifications}
\end{table}

\subsection{Networks}
\label{subsec:network}

\begin{figure*}
    \centering
    \includegraphics[width=\textwidth]{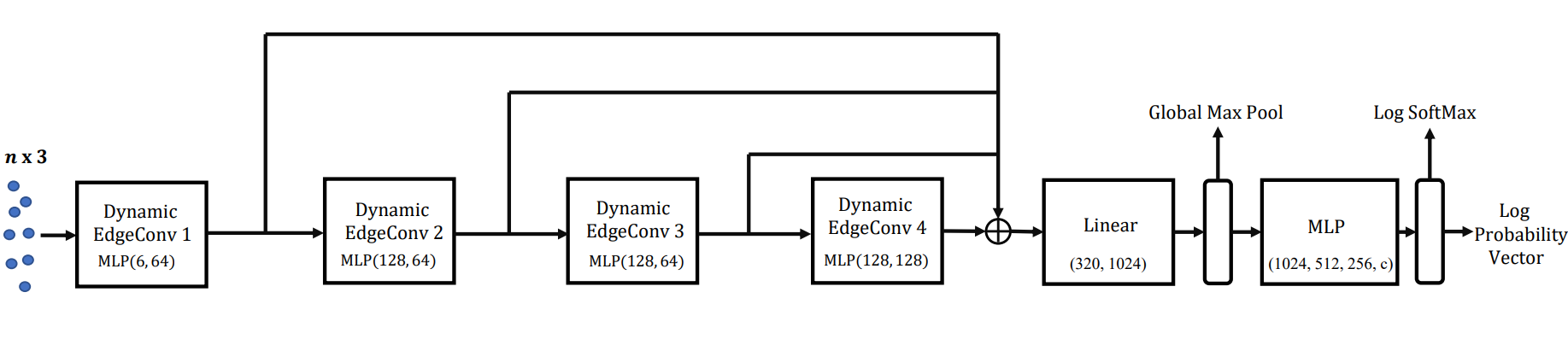}
    \caption{Network utilized in the experiment (for classification task). The input point cloud of shape $n \times 3$ ($n$ nodes with each node having $3$ features) is transformed by four successive Dynamic EdgeConv (DEC) layers. The Dynamic suffix underscores the graph construction that occurs in each EdgeConv layer via the $k$NN algorithm. The output of each DEC layer is concatenated to shape $n \times 320$, before being passed through a linear transformation that yields a $n \times 1024$ shaped output. A global max pool layer reduces this output to a vector of size $1024$ that is then passed through an MLP and a log-softmax layer to finally output the class log-probability vector.}
    \label{fig_3}
\end{figure*}

The base network that we utilize in the experiments is shown in Figure \ref{fig_3}. This network follows the network setting in \cite{wang2019dynamic} that achieves state-of-the-art results on the ModelNet40 \cite{wu20153d} graph-level classification dataset. The network comprises of four (dynamic) EdgeConv layers. Each layer constructs a $k$-nearest neighbor graph based on the current (latest) node embeddings before performing message passing on this graph via a single-layered $\mathcal{MLP}$ to update the embeddings. Each $\mathcal{MLP}$ in the EdgeConv layer utilizes a ReLU activation and includes a BatchNorm layer. The final $\mathcal{MLP}$ uses a dropout layer instead of BatchNorm. 

We train the above network for a range of $k$  values ($5$, $10$, $15$, $20$, $25$, $30$) for the EdgeConv layer. Additionally, we also train two quasi-Dynamic variants of such Dynamic GNNs by making static the last and the last two EdgeConv layers, respectively. Making static an EdgeConv layer, here, refers to removing the dynamic graph generating $k$NN block from an EdgeConv layer. In such quasi-DGNNs, we perform dynamic graph construction in each of the initial few EdgeConv layers of the network that form the dynamic portion of the network. The latter EdgeConv layers of the network (with the $k$NN block removed) do not reconstruct the graph again - this static portion of the network directly uses the last graph that was constructed in the dynamic portion of the network. We thus refer to dynamic EdgeConv layers (with $k$NN block) as Dynamic EdgeConv (DEC) and non-dynamic ones (without the $k$NN block) as simply EdgeConv (EC). Specifically, the last and the last two DEC layers of the network in Figure \ref{fig_3} are converted to EC layers to analyze the performance of quasi-DGNNs.

All the above networks are trained for a total of $100$ epochs on the entire ModelNet40 training dataset - we use the Adam optimizer \cite{kingma2017adam} with a learning rate of $0.001$ and a step learning rate scheduler with a gamma of $0.5$ and a step size of $20$. 

\subsection{Datasets}
\label{subsec:dataset}

We utilize the ModelNet40 \cite{wu20153d} dataset which contains $12,311$ graphs. We split the entire dataset into 80\% for training and 20\% for testing. The dataset comprises 3D point clouds in the form of CAD models from 40 categories - we pre-process the dataset by centering it and scaling it to a range of $(-1,1)$. Additionally, we sample $1024$ points during both training and testing. We test using a fixed random seed for reproducibility and equivalence across the tested networks.

\subsection{Performance Metrics}
\label{subsec:metrics}
We utilize classification accuracy as the performance metric. Since point cloud applications are usually real-time, we give importance to latency and memory consumption figures while assessing the networks performance. To this extent, we perform an exhaustive analysis of how latency and memory usage are distributed across a DGNN and within a DEC layer to identify valid performance trade-offs and bottlenecks.







\section{Results and Analysis}
\label{sec:result-and-analysis}

\subsection{Baseline Model Latency Analysis}
The baseline latency analysis is performed on a fully dynamic network (with all EdgeConv layers as Dynamic EdgeConv) (Figure \ref{fig_3}) with $k=20$. The value of $k$ is obtained by cross-validation over a set of values $(5, 10, 15, 20, 25, 30)$. For cross-validation, we split the training data into a train and validation set. Once a value of $k$ is selected, we re-train the network over the entire training data.

\begin{figure}[h!]
\centering

\begin{tabular}{cc}

  \includegraphics[width=0.23\textwidth]{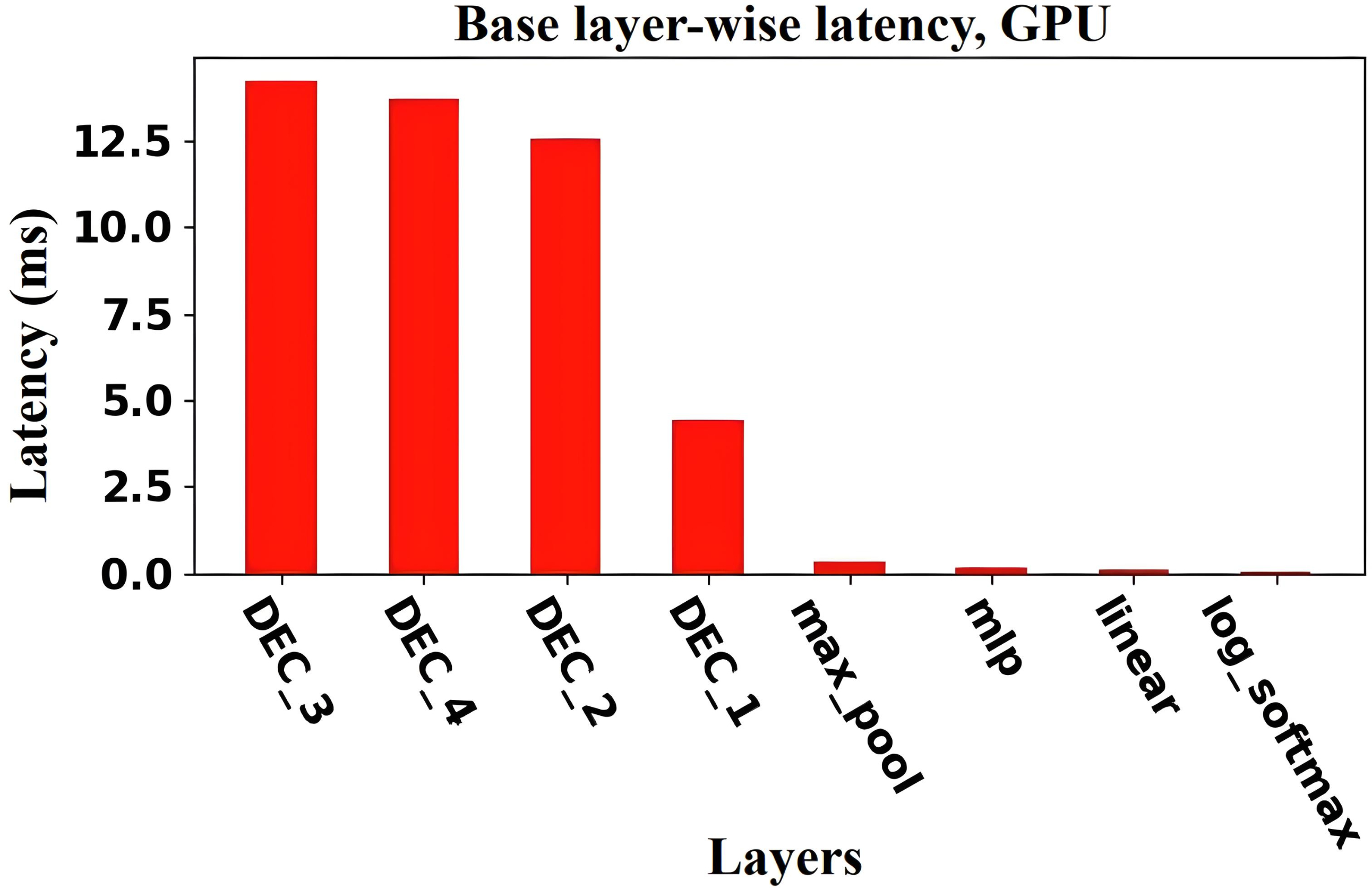} & \includegraphics[width=0.23\textwidth]{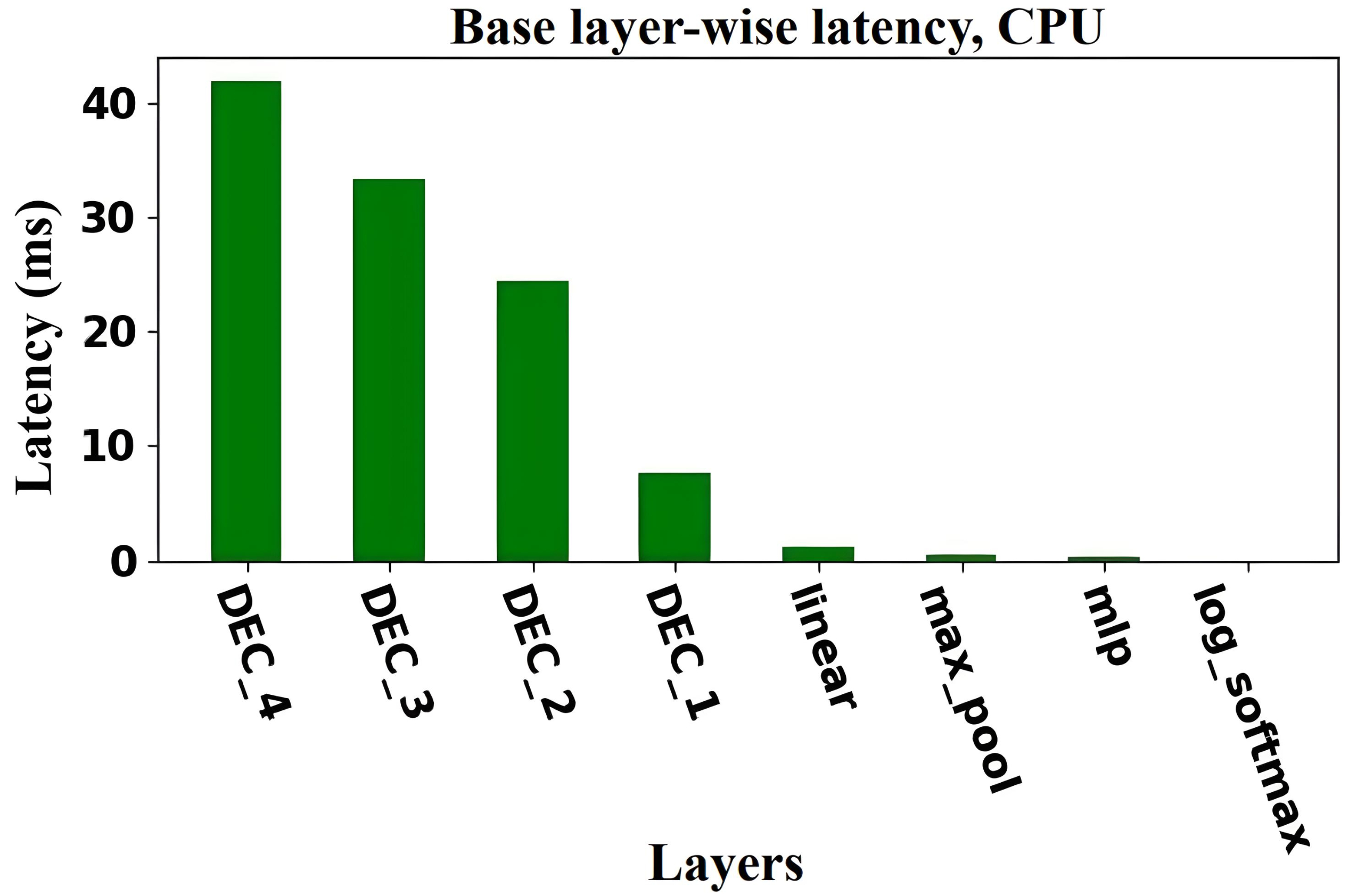} \\
  4(a)(i) &  4(a)(ii) \\
  [10pt]

  \includegraphics[width=0.23\textwidth]{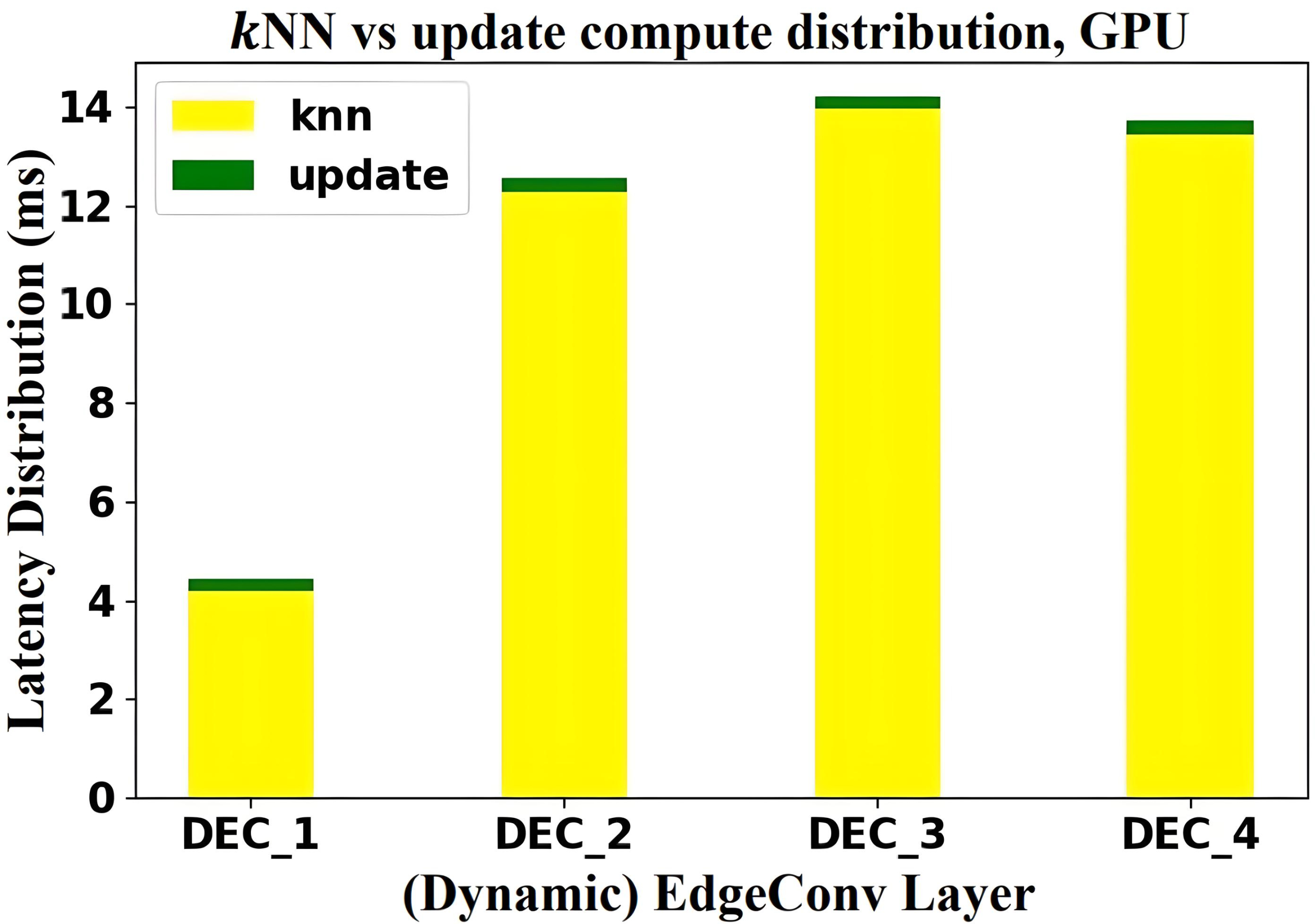} &   
  \includegraphics[width=0.23\textwidth]{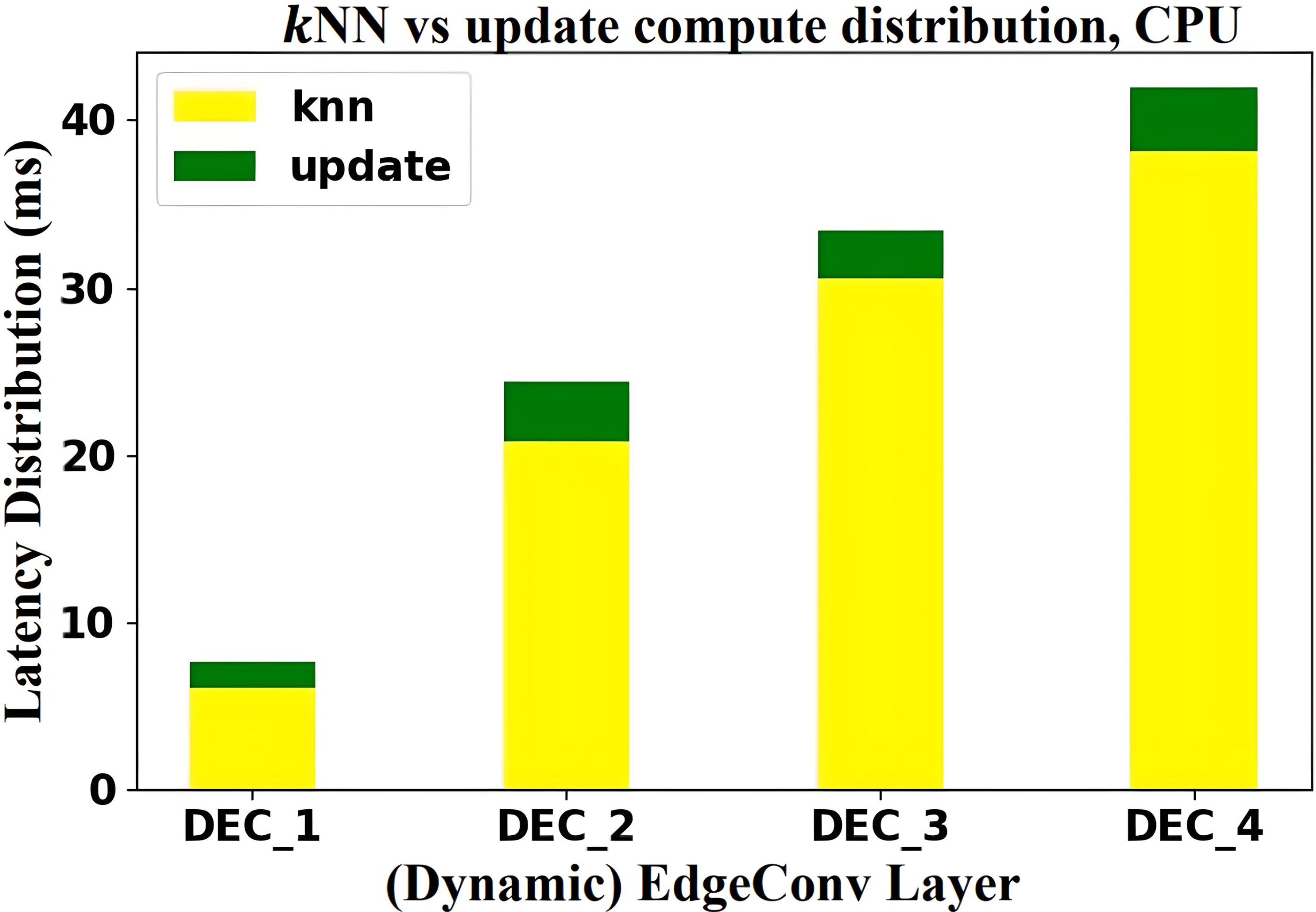} \\
  4(b)(i) & 4(b)(ii) \\
  [10pt]
  
\end{tabular}

\caption{Latency analysis of the baseline model. 4(a)(i) and 4(a)(ii) show the layer-wise latency (per-graph instance inference level) on GPU and CPU. DEC\_2, for example, indicates the second DEC layer in Figure \ref{fig_3}. 4(b)(i) and 4(b)(ii) analyze the individual DEC layers (comparing $k$NN vs update latency for each). Note that the EdgeConv layer refers to a DynamicEdgeConv layer.}
\label{fig:base_lat_anal}
\end{figure}

Figure \ref{fig:base_lat_anal} contains Figure 4(a)(i) and 4(a)(ii) that show the distribution of latency across all the layers of the network shown in Figure \ref{fig_3} on GPU and CPU, respectively.

Figures 4(b)(i) and 4(b)(ii) contain the per-layer latency analysis and latency distribution within the DEC layers for GPU and CPU, respectively. These figures indicate that the dynamic graph construction via the $k$NN algorithm is the bottleneck driving down the networks (and the DEC layers) performance.
\label{sec:latencybaseline}

\subsection{Analysis Under Varying $k$}
\label{sec:klatencyanalysis}
We analyze the effect of varying the number of nearest neighbors on the performance of the DEC layer and the point cloud classification model - we first train the network in Figure \ref{fig_3} for different values of $k$ associated with its DEC layers and then perform inference latency and accuracy analysis. As seen in Figure 5(a), the performance drops as we move away from the optimal $k$, this performance drop is sharper as we move towards the origin (towards smaller and smaller $k$'s). The network latency, for all $k$ values, on both CPU and GPU, is again dominated by the $k$NN algorithm.

\begin{figure}[h!]
\centering

\begin{tabular}{cc}

  \includegraphics[width=0.23\textwidth]{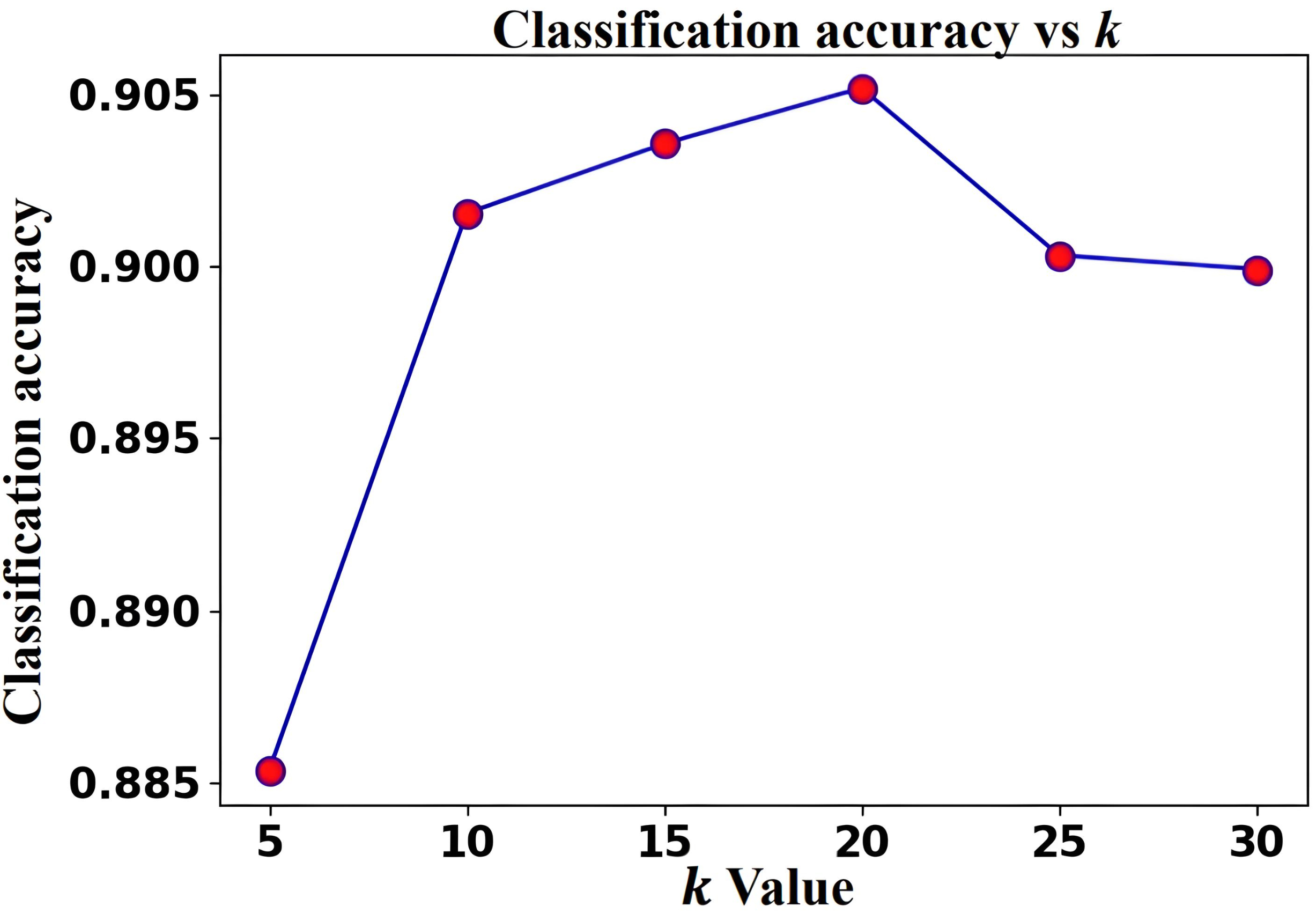} &   
  \includegraphics[width=0.23\textwidth]{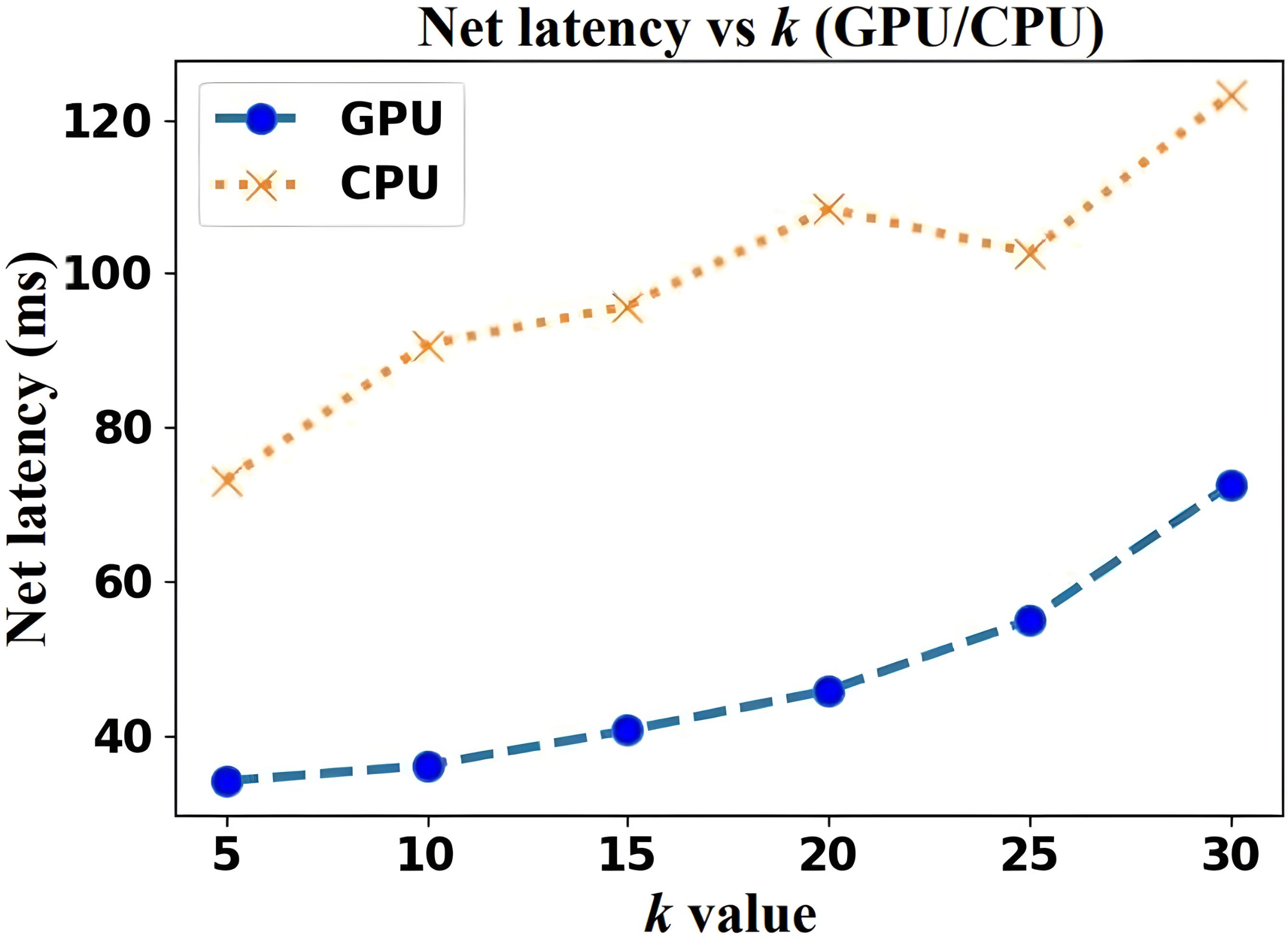} \\
  5(a) & 5(b) \\
  [10pt]

  \includegraphics[width=0.23\textwidth]{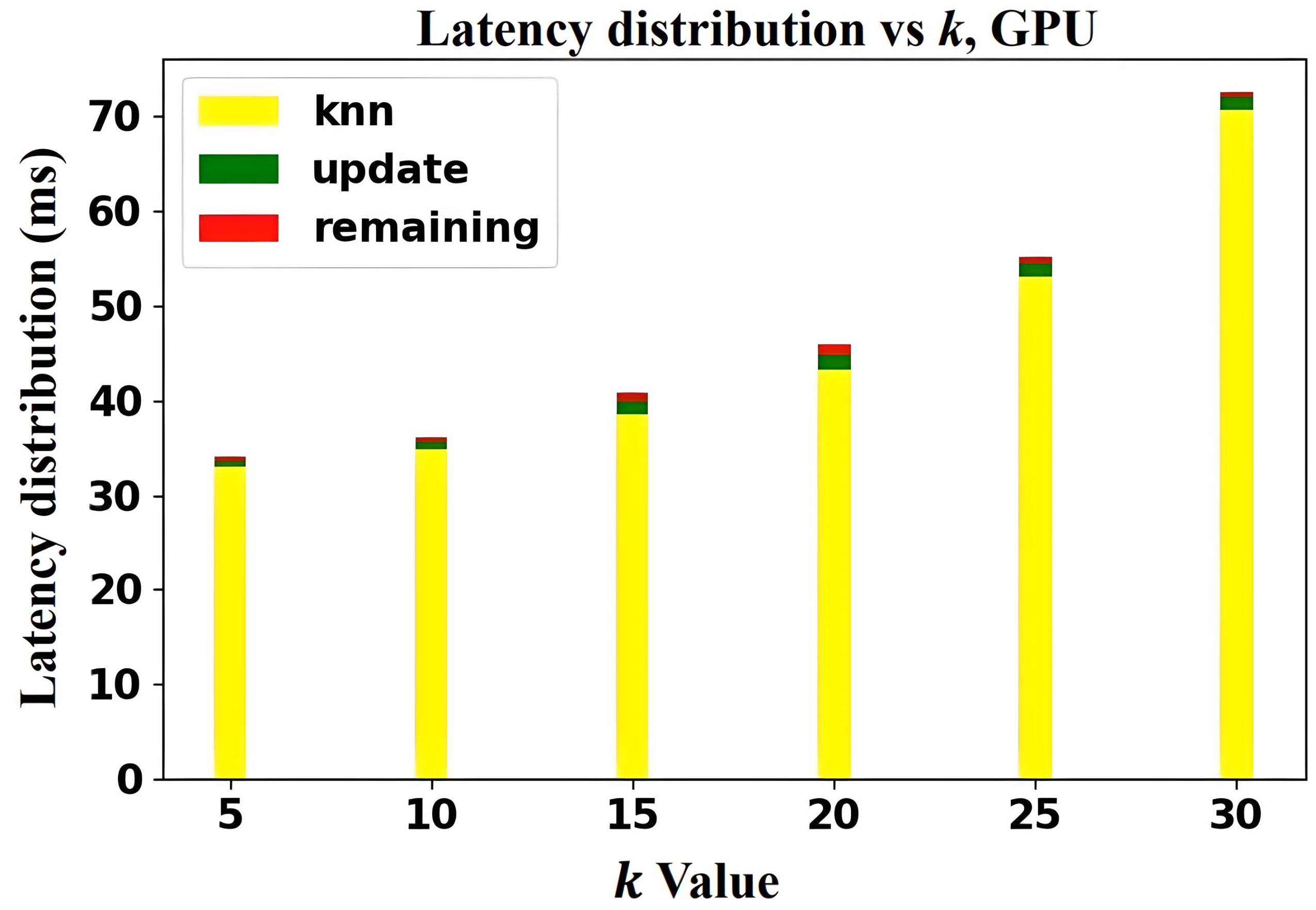} &   
  \includegraphics[width=0.23\textwidth]{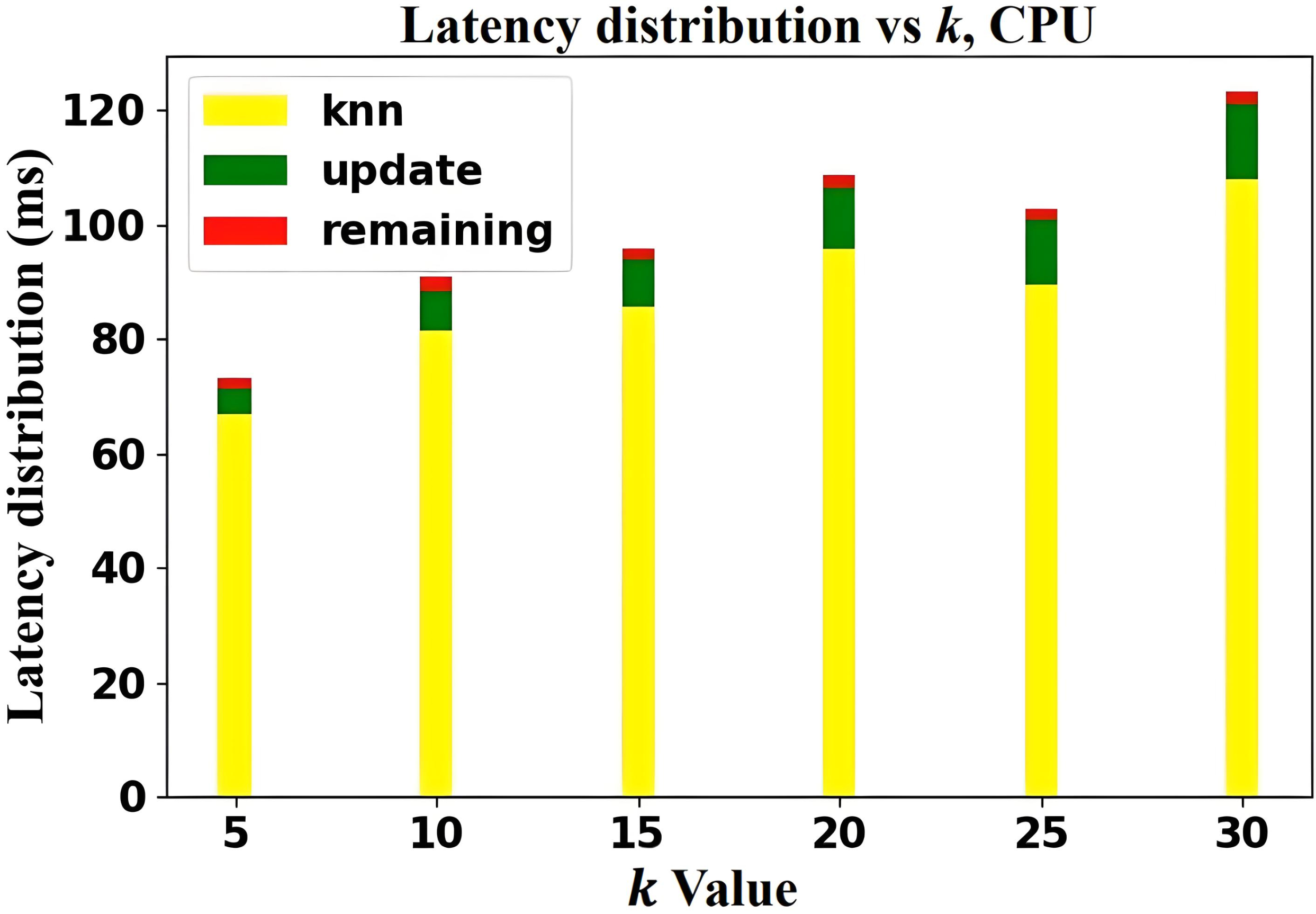} \\
  5(c)(i) & 5(c)(ii) \\
  [10pt]
\end{tabular}

\caption{Accuracy and latency analysis under varying $k$ values. 5(a) shows the variation of accuracy with $k$. 5(b), 5(c)(i) and 5(c)(ii) show the latency variation and distribution, respectively, versus $k$. The $update$ legend in 5(c)(i) and 5(c)(ii) is related to the message passing layer of the Dynamic EdgeConv layer.}

\end{figure}

\subsection{Quasi-Dynamic GNN (qDGNN)}
\label{sec:qDGNN}
Dynamic GNNs improve upon the performance of basic GNNs - this is markedly so for point cloud applications where DGNNs have the added capability of being able to identify and learn from semantically similar points irrespective of their geometric similarity (distance). Such DGNNs, however, as already seen, have a large computational cost linked to the dynamic graph construction operation which effectively bottlenecks the networks performance. 

Figure 6(a) clearly shows that making static the latter dynamic layers does not affect the networks accuracy - the corresponding performance gains associated with such Quasi-Dynamic GNNs are indicated in Figure 6(b), 6(c)(i) and 6(c)(ii) which show a drastic speed-up when compared against the fully dynamic baseline. This suggests that we can reach state-of-the-art performance, whilst being fast enough to operate at the edge, by utilizing a combination of dynamic and static EdgeConv layers for point clouds.

\begin{figure}[h!]
\centering

\begin{tabular}{cc}

  \includegraphics[width=0.23\textwidth]{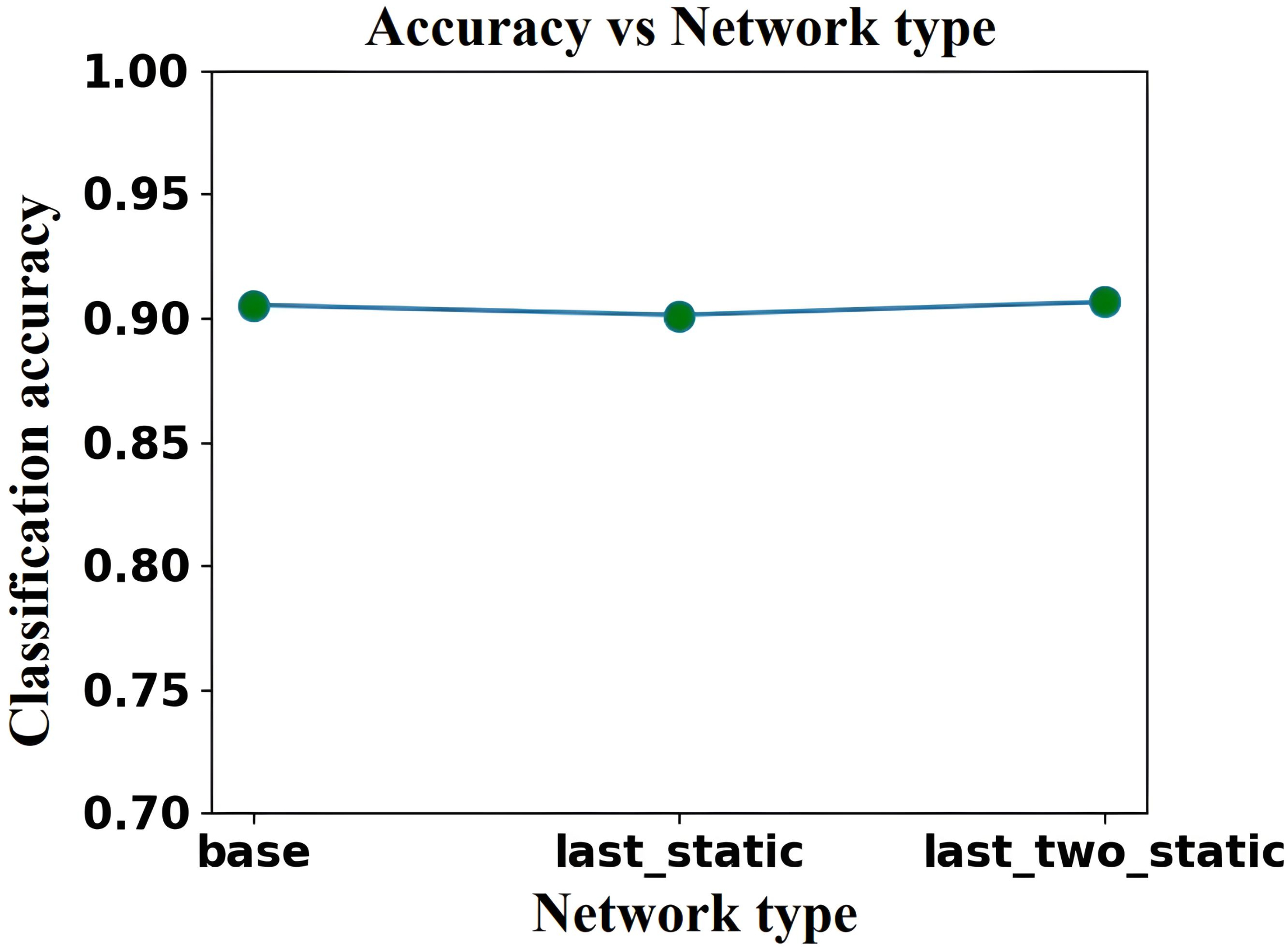} &   
  \includegraphics[width=0.23\textwidth]{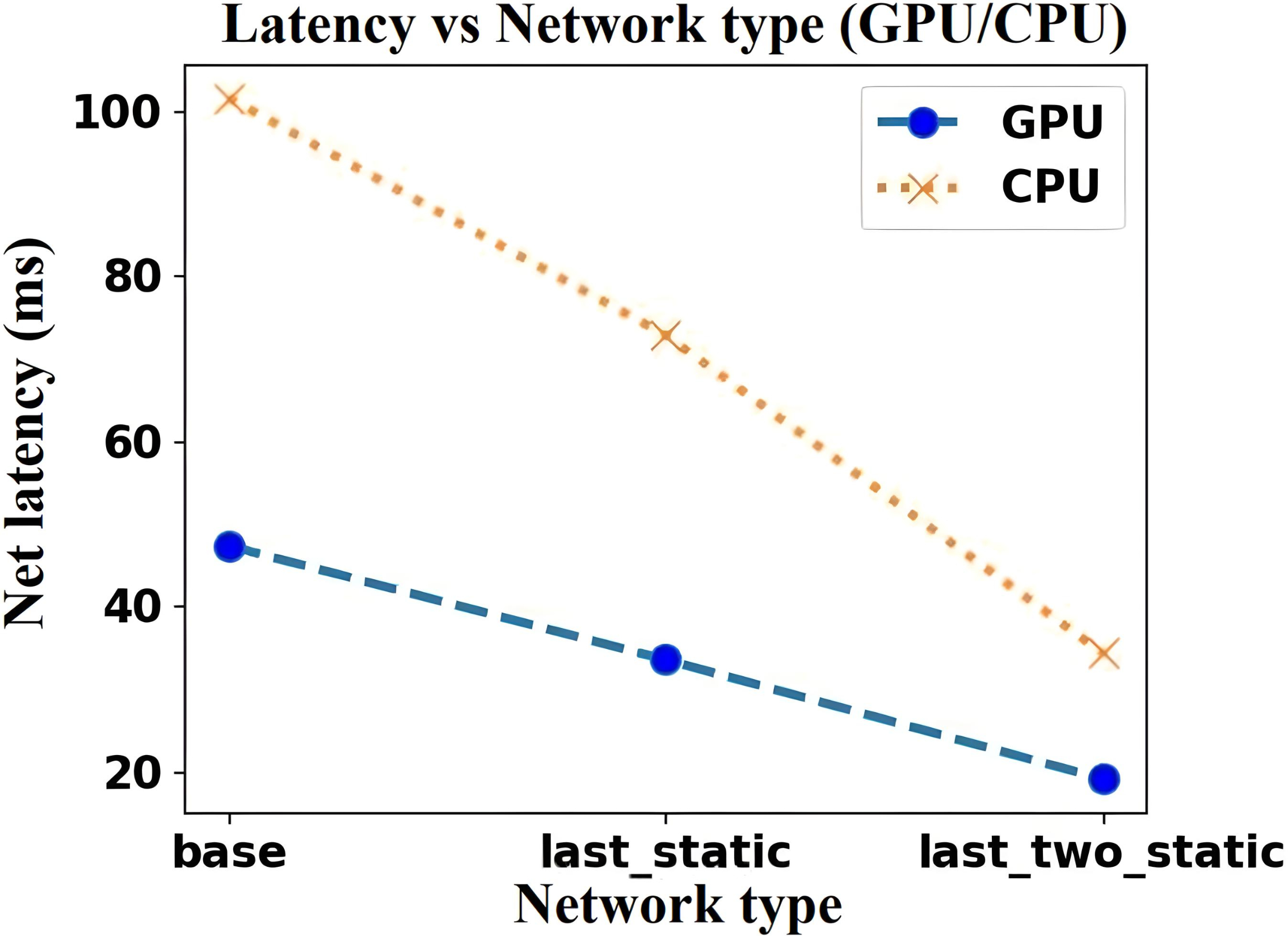} \\
  6(a) & 6(b) \\
  [10pt]

 \includegraphics[width=0.23\textwidth]{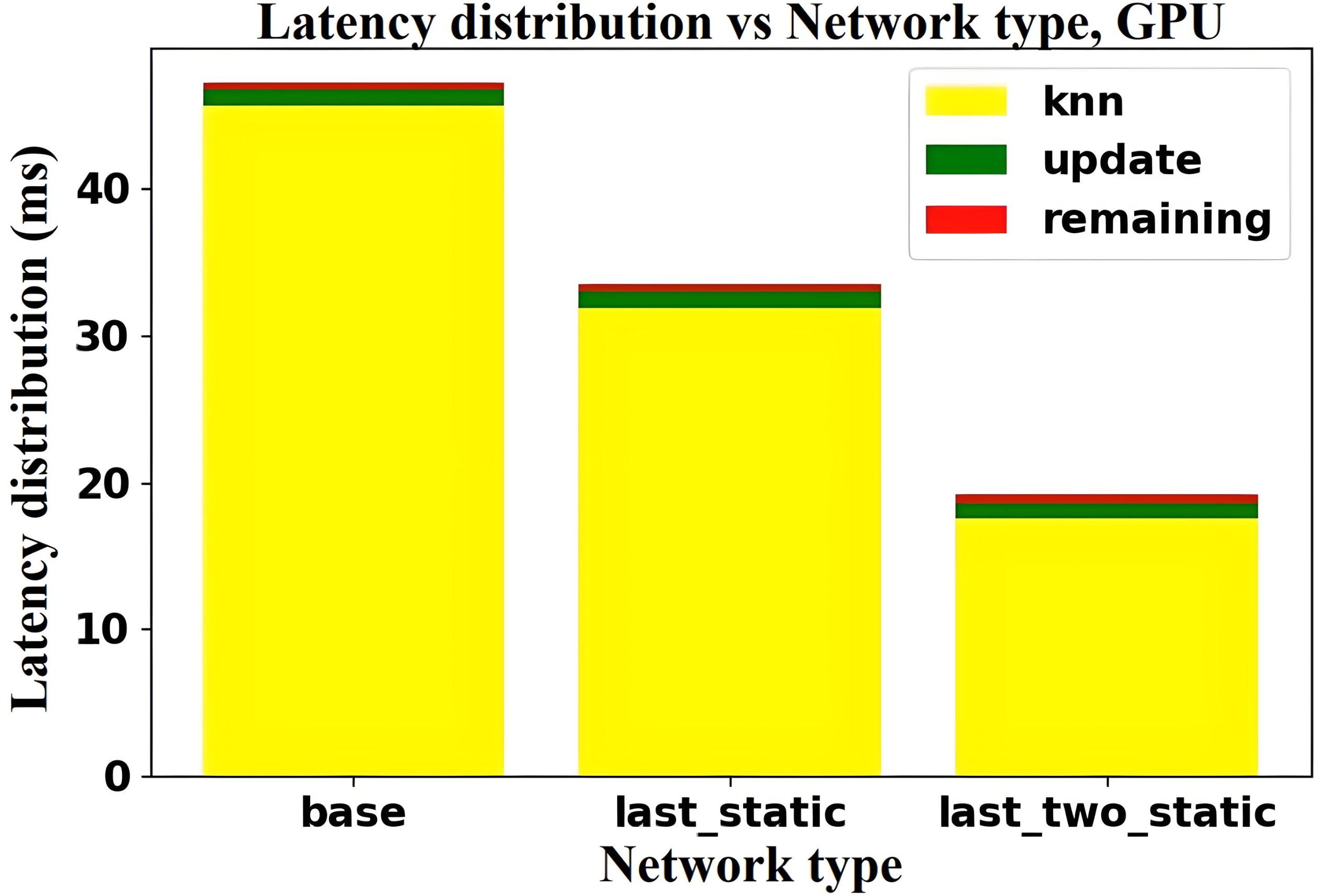} & \includegraphics[width=0.23\textwidth]{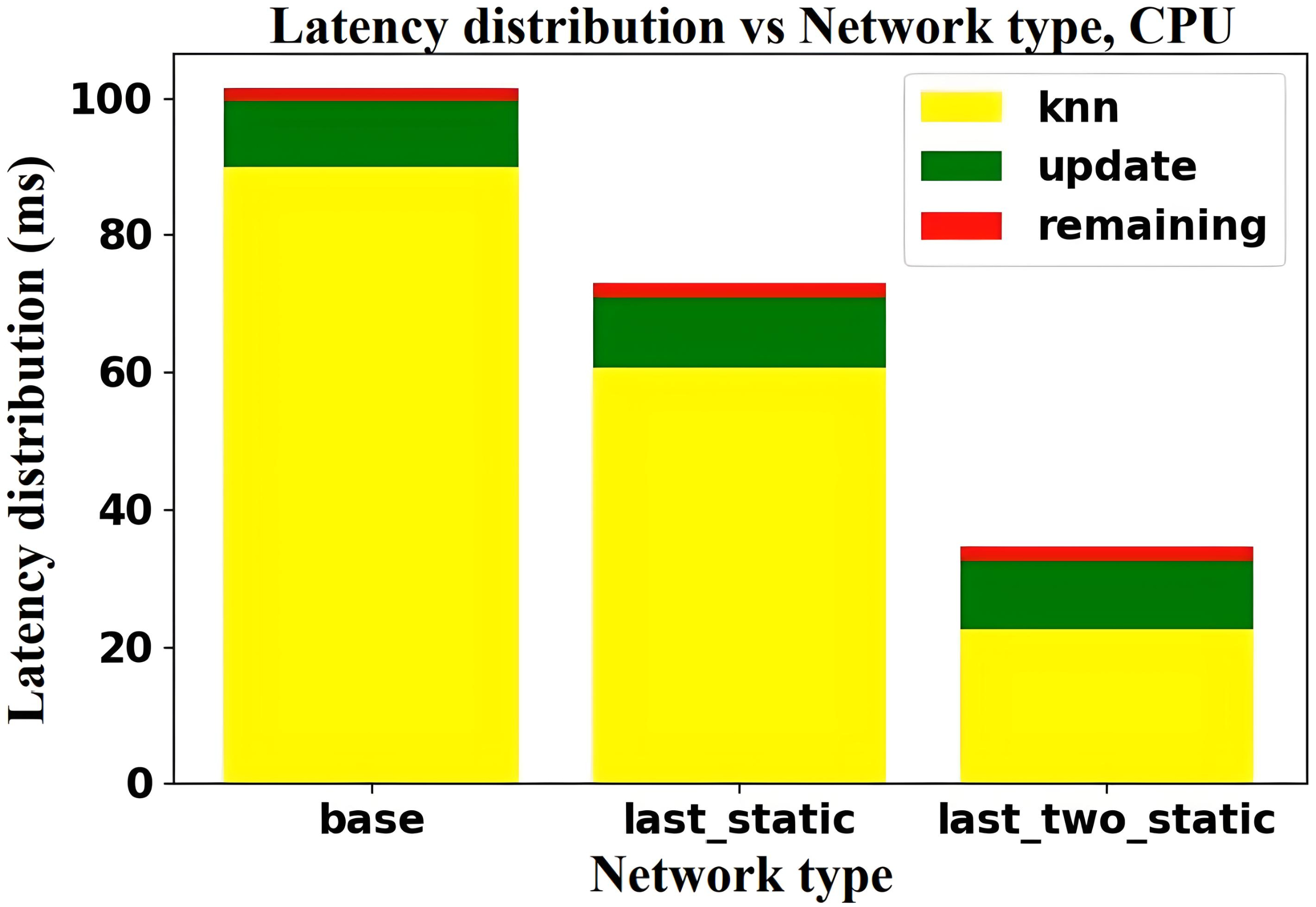} \\
 6(c)(i) & 6(c)(ii) \\
 [10pt]
\end{tabular}

\caption{Analysis of quasi-DGNN. 6(a) shows that qDGNN with last and last two layers static perform on-par with the fully dynamic baseline; 6(b), 6(c)(i) and 6(c)(ii) indicate that such qDGNNs are also much faster than their fully dynamic counterpart (refer Table \ref{sophisticatedtable}). Note that as we move away from the origin along the $x$-axis, our network becomes less dynamic (more static) (i.e. fewer DEC layers are employed).}

\end{figure}

\begin{figure}[h!]
\centering

\begin{tabular}{cc}

  \includegraphics[width=0.23\textwidth]{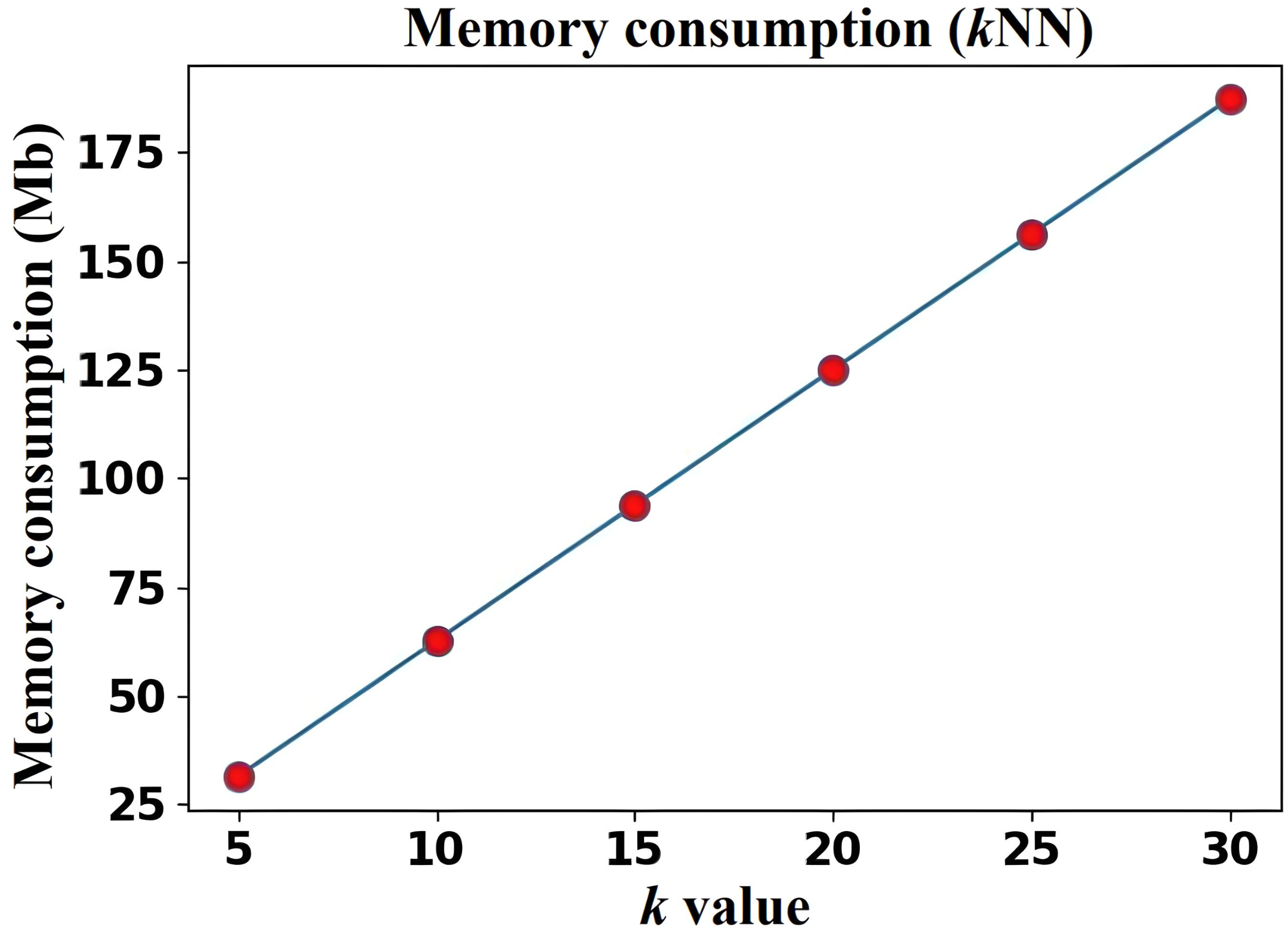} &   \includegraphics[width=0.23\textwidth]{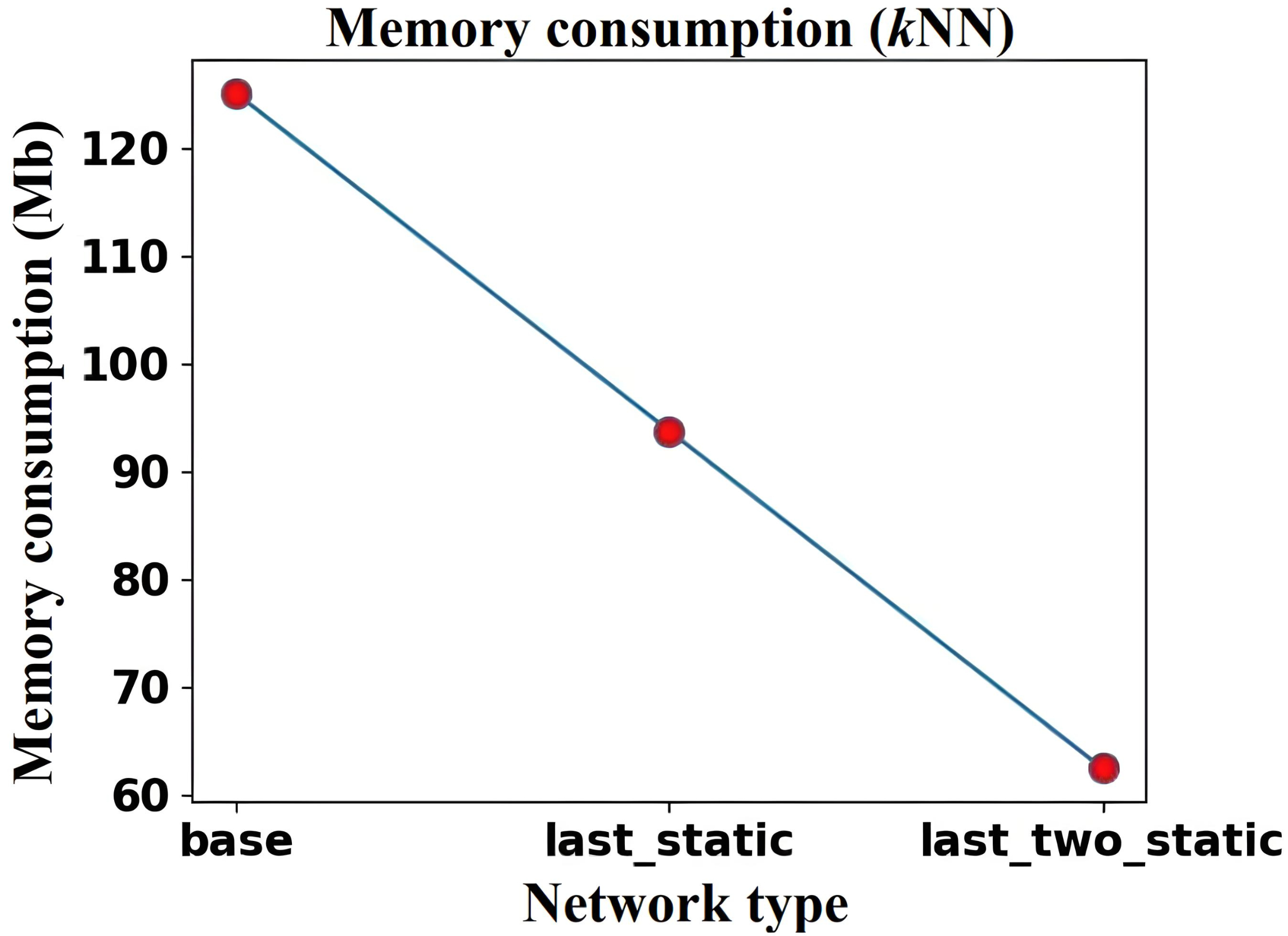} \\
  7(a) & 7(b) \\
  [10pt]
\end{tabular}

\caption{Memory consumption. 7(a) shows the memory consumption footprint of the $k$NN operation over $100$ single graph instance inferences for different values of $k$. 7(b) shows the same for different network types, last\_static and last\_two\_static refer to Figure \ref{fig_3} network but with the last and last two DEC layers converted to EC, respectively.}
\label{fig:mem_anal}
\end{figure}

\subsection{Memory Consumption}
\label{sec:mem}
As an example, the memory consumed for the $k$NN operation is plotted in Figure \ref{fig:mem_anal}; however, it is important to note that from a memory consumption point of view, the $k$NN graph construction operation is not a bottleneck - several other operators take up a much larger memory footprint compared to $k$NN. The linear nature of the curves is also self-explanatory, increase in $k$ leads to a proportional increase in memory required to serve the $k$NN operator. Meanwhile, removing $k$NN from latter layers of the network (DEC $\rightarrow$ EC) directly leads to a proportional reduction in memory consumption.

\setlength{\tabcolsep}{15pt}
\renewcommand{\arraystretch}{1.5}

\begingroup{}

\begin{table}[h!]
\captionof{table}{Latency and Accuracy}\label{sophisticatedtable}
\centering
\begin{tabular}{c|cc|c}
\hline
\multirow{2}{*}{Network (Figure \ref{fig_3})} & \multicolumn{2}{c|}{Latency (ms)}   & \multirow{2}{*}{Accuracy} \\ \cline{2-3}
                                  & \multicolumn{1}{c}{GPU}   & \multicolumn{1}{c|}{CPU}    &                           \\ \hline

$k$ = 5                             & \multicolumn{1}{c}{34.08} & 73.13  & 0.885                     \\

$k$ = 10                            & \multicolumn{1}{c}{36.13} & 90.76  & 0.901                     \\ 

$k$ = 15                            & \multicolumn{1}{c}{40.70} & 95.77  & 0.903                     \\ 

\boldsymbol{$k$} \textbf{= 20} \textbf{(baseline)}                     & \multicolumn{1}{c}{\textbf{45.49}} & \textbf{109.50} & \textbf{0.905}                     \\ 

$k$ = 25                            & \multicolumn{1}{c}{55.01} & 102.71 & 0.900                     \\ 

$k$ = 30                            & \multicolumn{1}{c}{72.58} & 123.28 & 0.899                     \\ 

Last Static                       & \multicolumn{1}{c}{33.52} & 72.83  & 0.901                     \\ 

Last Two Static                   & \multicolumn{1}{c}{19.12} & 34.37  & 0.906                     \\ \hline
\end{tabular}
\end{table}
\endgroup








\section{Discussion}
\label{sec:discussion}

The experimental results show the significant bottleneck introduced by the dynamic graph construction $k$NN layer in point cloud processing networks. $k$NN operation occupies up to $95\%$ latency of the (base) network on GPU and close to $90\%$ on CPU. Despite this, such a layer is crucial in boosting the network performance to enable a wide array of complex real-world edge applications. In this paper, we shed light on this problem whilst also providing a simple solution - quasi-Dynamic Graph Neural Networks (qDGNN). Such networks significantly improve the latency of the network (reduction in latency by up to $58\%$ on GPU and up to $69\%$ on CPU) whilst maintaining the same level of performance as is demonstrated by DGNNs. Accelerating $k$NN layers on FPGA and deploying the DGNN on an FPGA platform is also a potential solution that has attracted significant research interest recently \cite{auten2020hardware, 8594633}; optimizing $k$NN algorithms \cite{dong2011efficient} is also an area that has been studied with much interest.


\section{Conclusions and Future Work}
\label{sec:conclusions}

In this paper, we examined the latency bottleneck associated with a DynamicEdgeConv (EdgeConv) layer whilst examining its inference performance - optimizing the dynamic graph construction stage in such networks is a problem that remains to be solved and a promising research avenue due to its scope of applicability.




\section*{Acknowledgement and Statement}
This work is supported by the  DEVCOM  Army Research Lab (ARL) under grant W911NF2220159.  

\textbf{Distribution Statement A}: Approved for public release. Distribution is unlimited. 

\bibliographystyle{IEEEtran}
\bibliography{ref}

\end{document}